\documentclass[aps,prd,onecolumn,groupedaddress,showpacs,nofootinbib,amssymb]{revtex4}
\usepackage[dvips]{graphicx}
\usepackage{amssymb}
\usepackage{amsmath}
\usepackage{graphicx,,color}
\usepackage{amsfonts}
\usepackage{bm}
\usepackage{cancel}
\usepackage{comment}

\newcommand\be{\begin{equation}}
\newcommand\ee{\end{equation}}

\newcommand{\G}{{\mathcal G}}
\allowdisplaybreaks[4]

\begin{document}

\tolerance=5000

\title{  Geometric Perfect Fluids from Extended Gravity}
\author{Salvatore Capozziello$^{1,2,3}$, \thanks{capozziello@na.infn.it} Carlo Alberto Mantica$^{4}$, \thanks{carlo.mantica@mi.infn.it}  Luca Guido Molinari$^{4}$ \thanks{Luca.Molinari@unimi.it}}
\affiliation{$^{1}$ Dipartimento di Fisica, ''E. Pancini'' Universit\`a
''Federico II'' di Napoli, \\and
INFN Sez. di Napoli, Compl. Univ. Monte S. Angelo Ed. G,
Via Cinthia, I-80126 Napoli, Italy,
\\
$^{2}$ Scuola Superiore Meridionale, Largo S. Marcellino 10, I-80138 Napoli, Italy,\\
$^{3}$ Laboratory for Theoretical Cosmology, International Center of Gravity and Cosmos, Tomsk State
University of Control Systems and Radioelectronics (TUSUR), 634050 Tomsk,
Russia,\\
$^{4}$Physics Department ``Aldo Pontremoli'',
Universit\`a degli Studi di Milano\\ and INFN sez. di Milano,
Via Celoria 16, 20133 Milano, Italy.
}

\date{\today}
\tolerance=5000

\begin{abstract}
A main issue in cosmology and astrophysics is whether the dark sector phenomenology originates from particle physics, then requiring the detection of new fundamental components,
or it can be addressed by modifying General Relativity. Extended Theories of Gravity are possible candidates aimed in framing dark energy and dark matter in a comprehensive geometric view.
Considering the concept of perfect scalars, we show that the field equations of such theories naturally  contain perfect fluid terms. Specific examples are developed for the Friedman-Lema\^itre-Roberson-Walker metric.
\end{abstract}

\pacs{04.50.Kd, 95.36.+x, 98.80.-k, 98.80.Cq}

\maketitle

\section{Introduction}
Since Hubble disclosed the realm of the nebulae, the Cosmological Principle has been confirmed by a vast range of observations, and it is the 
cornerstone of the Cosmological Standard Model. The large scale uniformity and isotropy of space at all times fixes the Friedman-Lema\^itre-Robertson-Walker (FLRW) metric on purely geometrical grounds.  However, the evolution of cosmological  parameters is given by  the Einstein field equations once the sources are specified.
Besides the  perfect fluid matter sources,  such  equations  must account for observed gravitational effects that are now ascribed to  {\em dark energy} and {\em dark matter} whose effects are well-evident at astrophysical and cosmological scales. 

According to the basic picture, 
in the framework of the $\Lambda$CDM model, dark energy behavior  can be addressed by  the cosmological term $\Lambda g_{\mu\nu}$. 
Dark matter is a dust term added to the energy-momentum tensor of ordinary matter and describes, in the standard expectation, a sea of unknown weakly interacting particles. At higher resolution, in response to primordial perturbations, it constitutes the cosmic web where galaxy clusters form.

There is a huge global effort to detect such dark particles, which are probably strangers to the standard table of elementary particles, but,  at the moment, there is no univocally determined particle candidate capable of matching consistently the whole phenomenology despite  the efforts of extremely precise running experiments \cite{Aleksandrov}.

But new particles are not necessarily the correct answer. Dark sector could be nothing else but the indication that General Relativity should be revised or extended at infrared scales so the answer to trace a consistent cosmic history could be of geometric origin \cite{Capoz02,Bamba}.

The key point for this program is to show that any further geometric term beyond the Ricci curvature scalar $R$ or any other geometric invariant in the gravitational action can be model out as a {\it perfect fluid}. This result should be not only an analogy but a direct demonstration. In such a case, within the paradigm of General Relativity,  a correct picture of cosmic flow  could definitely address  the problem of dark matter and dark energy. 

 The purpose of this perspective paper is to confirm the broad consensus that  dark effects may be searched in  geometry, by extending the Hilbert-Einstein action  from its simplest original form. We shall consider some well known models of extended gravity: $f(R)$, $f(R,\G)$, Higher-order, Quadratic gravity. For all of them, adopting the notion of \textit{perfect scalar}, we show that  perfect fluid terms appear in the field equations. We specialize this analysis to the case of FLRW space-time but it can be enlarged to more general space-times.
 In this perspective, data can be used to reconstruct the "correct" cosmic history without searching for new particle candidates \cite{Odi2}. Clearly, initial ultraviolet behavior has to be considered and matched with the late time epochs and large scale structure  in a unitary picture \cite{Odi1}.
 
The layout of the paper is the following. In Sec. II, we define the perfect scalars in the context of FLRW spacetimes. 
This concept allows to deal with any Extended Gravity under the standard of perfect fluids as discussed in Sec.III, IV, V, and VI where relevant examples are given. Conclusions are drawn in Sec. VII.

\section{Perfect Scalars in Friedman-Lema\^itre-Robertson-Walker spacetimes}

The Cosmological Principle fixes the structure of spacetime as a warped manifold, where time-slices are Riemannian subspaces of constant curvature \cite{Wald}.
In proper coordinates the FLRW metric is:
\begin{equation}
ds^2 = -dt^2 + a^2(t) \left [ \frac{dr^2}{1- k r^2} + r^2 d\Omega \right ], \quad k=0,\pm 1 \label{RWm}
\end{equation}
In a covariant setting\footnote{We adopt the notation: $i,j,k,..=0,1,2,3$; $\mu,\nu,\rho, ...=1,2,3$. A dot means the operator  $u^k\nabla_k $, i.e. the time derivative in the comoving frame $u^0=1$, $u^\mu=0$.}, a FLRW spacetime is characterized by a time-like unit vector field, $u_k u^k=-1$, that is shear-free, vorticity-free
and acceleration-free, which means
\begin{eqnarray}
&&\nabla_j u_k = {\sf H} (g_{jk}+u_j u_k)\,, \label{torq} \\
&& \nabla_j {\sf H} = -u_j \dot {\sf H}\,,
\end{eqnarray} 
and by a null Weyl tensor, $C_{jklm}=0$ (see also \cite{MaMoJMP16} where the Chen vector $X_j\propto u_j$ is used).   The tensor
$h_{jk}=g_{jk}+u_ju_k$ is the projection on the vector space locally orthogonal to $u^k$. 
With these properties and the relation $R_{jklm}u^m =(\nabla_j\nabla_k -\nabla_k\nabla_j)u_l$ one obtains
the Riemann tensor \cite{MaMoJMP16}:
\begin{eqnarray}
R_{jklm} =&& \tfrac{1}{6} (R-2\xi) (g_{km} g_{jl} - g_{kl}g_{jm}) \label{Riemann}\\
&& + \tfrac{1}{6}(R-4\xi) (g_{km} u_j u_l - g_{kl} u_j u_m  -g_{jm} u_ku_l \,,
 + g_{jl} u_k u_m)  \nonumber
\end{eqnarray}
and the Ricci tensor,  with the perfect fluid form, 
\begin{equation} 
R_{kl} = \frac{1}{3}(R-4\xi) u_ku_l  + \frac{1}{3}(R-\xi) g_{kl}  \label{Ricci}
\end{equation}
where $\xi $ is the eigenvalue: $R_{jk}u^k =\xi u_j$,
\begin{equation}
\xi = 3({\sf H}^2 + \dot {\sf H}).
\end{equation}
In the comoving frame \eqref{RWm}, Eq.\eqref{torq} shows that the field ${\sf H} $ is precisely the Hubble parameter 
and $\xi $ is related to the deceleration parameter $q=-\ddot a a/\dot a^2$:
\begin{equation}
{\sf H}(t) = \frac{\dot a}{a}, \qquad \xi = 3 \frac{\ddot a}{a} =-3{\sf H}^2(t) q(t)\,. 
\end{equation}
For a detailed discussion see \cite{Boskoff}.

The following definitions and lemmas have a key role in the forthcoming discussion of cosmological models derived from  Extended Gravity \cite{CMM20HO}. According to their validity, it is possible to model out further geometric terms in the Hilbert-Einstein action as perfect fluids. Let us start from the definition:
\begin{center}
$$\boxed{
\mbox{A scalar field $S$ is {\em perfect} if $\nabla_i S= -u_i \dot S$.}}$$
\end{center}

\noindent \textbf{Lemma}  \label{lem1} \\
\textit{If $S$ is perfect, then $\dot S$ is perfect, i.e. $\nabla_j \dot S =-u_j \ddot S$.}
\\

\noindent \textbf{Proof}  \\
$\nabla_i \dot S = \nabla_i (u_k\nabla^k S) = {\sf H} h_{ik} \nabla^k S  + u^k\nabla_i\nabla_k S$. The first term is zero
because $S$ is perfect. In the second term, let us  exchange derivatives and obtain $u^k\nabla_k (\nabla_i S) =
-u^k\nabla_k (u_i\dot S) = -u_i u^k\nabla_k \dot S$. Then: $\nabla_i \dot S = -u_i \ddot S$ i.e. $\dot S$ is perfect.$\Box$
\\ 
\\
If $S$ is perfect then $f(S)$ is perfect for a smooth function. Sums and products of perfect scalars are perfect.
Since ${\sf H} $ is perfect, then ${\sf H}^2$ and $\dot{\sf H} $ are perfect, and then 
$\xi $ is perfect.
\\

\noindent \textbf{Lemma}  \label{lem2}\\
\textit{If $S$ is perfect then the Hessian $\nabla_i\nabla_j S$ has the perfect-fluid form:}
\begin{equation}
\nabla_i\nabla_j S= -{\sf H}  \dot S g_{ij} +(\ddot S -{\sf H} \dot S) u_iu_j \label{Spf}
\end{equation}

 \noindent \textbf{Proof}  \\
$$\nabla_i\nabla_j S = \nabla_i(-u_j \dot S) = -{\sf H} h_{ij} \dot S -u_j \nabla_i \dot S= -{\sf H} h_{ij} \dot S +u_iu_j 
\ddot S.$$
$\Box$\\ 

In particular, if $S$ is perfect, then $\square S = -3{\sf H} \dot S -\ddot S $ is perfect, being a combination of 
perfect fields, and the scalar fields $\square^m S$, $m=1,2,...$  are perfect.
\\

\noindent\textbf{Lemma} \\
\textit{$R$ is a perfect scalar, $\nabla_j R=-u_j\dot R$, and }
\begin{equation}
\dot R - 2\dot\xi = - 2{\sf H} (R-4\xi) \label{dotR} 
\end{equation}
\\

\noindent\textbf{Proof}  \\
The Bianchi identity $\nabla^k R_{kj}=\frac{1}{2}\nabla_j R$ is evaluated with \eqref{Ricci}. That is:
\begin{eqnarray*}
 \tfrac{3}{2}\nabla_j R &=& \nabla^k [(R-4\xi) h_{kj}] +3\nabla_j\xi\\
  &=&\nabla_j R -\nabla_j \xi + u_j(\dot R -4\dot \xi )+ 3(R-4\xi)  {\sf H} u_j \\
 & =& \nabla_j R +u_j \dot R +3u_j [{\sf H} (R-4\xi) -\dot\xi ]
\end{eqnarray*}
Contraction with $u^j$ gives \eqref{dotR} and  the last equation gives $\nabla_j R =-u_j \dot R$. $\Box$
\\ 
\\
The differential Eq. \eqref{dotR} can be  solved in the comoving frame. Multiply it by $a(t)$, one  obtains:
$\displaystyle{\frac{d}{dt}(Ra^2 -2 \xi a^2) = 4\xi \dot a a}$. Insert $\xi =3\ddot a/a$ and obtain the total derivative:
$\displaystyle{ \frac{d}{dt}(R a^2- 6 a\ddot a - 6 \dot a^2)=0}$. Integration yields
\begin{equation}
R = \frac{R^*}{a^2} +  6\left(\frac{\ddot a}{a}\right) + 6\left(\frac{\dot a}{a}\right)^2 \label{RRstar}
\end{equation}
where the number $R^*$ is the curvature scalar of the space submanifold. By rescaling space coordinates, it can be put equal to $k=0,\pm 1$  (see \cite{Boskoff}).
This equation is independent of the model and of the matter content of the universe, and only descends from the FLRW hypothesis that the space-time is homogeneous and isotropic. 

In Ref.~\cite{MolMan19a} the simplest choice $\dot R=0$ and $k=0$ yields the expanding solution $a^2(t) = \sinh (t \sqrt{R/3} )$. This {\em a priori} choice of the 
geometry has an interesting outcome: the 
Einstein equations with $\Lambda =0$ imply an evolving perfect fluid with $p=\rho /3$ at $t\ll \sqrt R $ and $p=-\rho $ at $t\gg \sqrt R$. This means that a transition from a matter dominated phase and an accelerated behavior can be achieved according to the value of the field $\sqrt R$.

In real world, there is a complex interplay between geometry and matter. 
The evolution of  $a(t)$ is determined by non-linear, model-dependent, generalized Friedmann equations. 
As we will see below,  the concept of perfect scalar is useful to define  fluids sourcing the cosmological equations, independently of their matter or geometric origin \cite{CMM20HO}. Specifically, the Hessian of a perfect scalar has the perfect fluid form. 
Since the Ricci scalar is the fundamental geometric object, the property of being {\it perfect} requires that the Weyl tensor is harmonic. 

In next sections, we consider different Extended Theories of Gravity and show that the field equations only contain perfect fluid terms. In other words, according to the previous prescriptions, any geometric contribution can be recast as a perfect fluid.

\section{$f(R)$ gravity}
$f(R)$ theories of gravity were introduced by Buchdahl in 1970 \cite{Buchdahl70} and gained popularity with the works by Starobinsky on cosmic inflation \cite{Starob80}. Now
they are explored as possible theories to avoid the need of dark matter and dark energy \cite{CF08,Faraoni,CDL11, Lambiase}. 
In these theories, the scalar $R$ in the gravitational 
action is replaced by a smooth function $f(R)$: 
\begin{equation*} 
S = \frac{1}{2\kappa} \int d^4 x \sqrt{-g} f(R) + S^{(m)}
\end{equation*}
$S^{(m)}$ is the action for the matter fields. The vanishing of the first variation in $g_{kl}$ gives, modulo surface terms, the field equations:
\begin{eqnarray} 
&& f'(R)R_{kl}  - [f^{'''}(R) (\nabla_k R)(\nabla_l R) + f^{''}(R) \nabla_k\nabla_l R] \label{fieldeqs2}\\
&& + g_{kl} [ f^{'''}(R) (\nabla_k R)^2 + f^{''}(R) \square R-\tfrac{1}{2} f(R)]=
\kappa T_{kl} \nonumber
\end{eqnarray}
where a prime denotes a derivative with respect to $R$. 
The contracted Bianchi identity $\nabla_k T^k{}_l=0$ is preserved for any smooth $f(R)$.

By describing ordinary matter as a mixture of perfect fluids with pressure $p_m$ and energy density $\rho_m$, 
the left hand side must have the form of a perfect fluid. 
In FLRW spacetimes, the Ricci tensor is a perfect fluid, according to Eq.\eqref{Ricci}, and we showed that $R,\xi$ are perfect scalars.
Then we obtain the perfect fluid expression
\begin{eqnarray*} 
&& \tfrac{1}{3}f'(R)[(R-4\xi)u_ku_l +(R-\xi)g_{kl}]  - f^{'''}(R)  \dot R^2 u_k u_l+ f^{''}(R) [{\sf H} \dot R h_{kl}  - u_k u_l \ddot R] \\
&& + g_{kl} [ - f^{'''}(R) \dot R^2 - f^{''}(R) (3{\sf H} \dot R + \ddot R) -\tfrac{1}{2} f(R)]=
\kappa (\rho_m u_ku_l + p_m h_{kl}).\nonumber
\end{eqnarray*}
Separation of terms proportional to $h_{kl}$ and to $u_ku_l$ gives:
\begin{eqnarray*} 
&& \tfrac{1}{2} f(R)- \tfrac{1}{3}f'(R)(R-\xi) + f^{''}(R) (2{\sf H} \dot R + \ddot R)  +f^{'''}(R) \dot R^2  = -
\kappa p_m\\
&& \tfrac{1}{2} f(R)- f'(R)\xi  + 3f^{''}(R) {\sf H}  \dot R    =\kappa \rho_m
\end{eqnarray*} 
For $f(R)=R$ we recover the General Relativity and then $ R + 2\xi  = -6\kappa p_m$,  $R- 2\xi    = 2\kappa \rho_m$.  
The extra terms coming from $f(R)-R$ are an effective pressure and energy density of geometric origin, that add to 
the ordinary matter terms. General properties of vacuum solutions $p_m=0$ and $\rho_m =0$ are discussed in \cite{CMM20vac} .

In the Starobinski model, $f(R)=R+\alpha R^2$, the equations become
\begin{eqnarray*} 
&& \tfrac{1}{2} (R+2\xi) -\tfrac{1}{2}\alpha (R^2 -4R\xi - 24{\sf H} \dot R -12 \ddot R)    = -3\kappa p_m\\
&& \tfrac{1}{2}(R-2\xi) +\tfrac{1}{2}\alpha (R^2 - 4 R\xi  + 12 {\sf H}  \dot R)   = \kappa \rho_m
\end{eqnarray*} 
In flat space, Eq.\eqref{RRstar}, with $R^*=0$, gives a non-linear fourth-order equation for $a(t)$.

In Ref.~\cite{CMM19fR} we proved the following statement: the sufficient conditions for a quasi-Einstein spacetime 
(i.e. with perfect fluid Ricci tensor) to separately have $\nabla_i R \nabla_j R$ and $\nabla_i\nabla_j R$ of perfect fluid form are: 
\begin{enumerate}
\item  the spacetime is: $ds^2=-dt^2+a^2(t)g^*_{\mu\nu} dx^\mu dx^\nu$, 
that is the space can be also non-homogeneous and anisotropic, more general than FLRW space-time;
\item the Weyl tensor has to be  harmonic ($\nabla_m C_{jkl}{}^m=0$).
\end{enumerate}
If they hold,   field Eqs.\eqref{fieldeqs2} describe perfect fluids. In other words, in the case of $f(R)$ gravity, the extra geometric terms give rise to a further perfect fluid with respect to the standard matter one.

In fact, according to the above definition,  $f(R)$ gravity has one more perfect scalar compared to GR. By a rapid inspection of the field equations, such a perfect scalar  is connected to the tensor degrees of freedom.
In this context, the possible detection of  further polarization modes  of gravitational waves is a crucial issue to define the number of degrees of freedom and then to select a self-consistent theory of gravity \cite{Bobby}.

\section{Gauss-Bonnet gravity}
An important extension of General Relativity is related to the Gauss-Bonnet scalar  
\begin{equation}
\G= R_{jklm}R^{jklm} -4 R_{jk} R^{jk}+ R^2\,, \label{GB}
\end{equation}
which is a topological invariant related to the Euler characteristic of a manifold.
Nojiri and Odintsov \cite{NojOd05} introduced modified gravity in the form $R+f(\G)$ as an alternative to $f(R)$ to solve, on geometric grounds,
the problem of dark energy and the late acceleration of the universe. Besides, the Gauss-Bonnet term is useful in regularizing  quantum fields in curved spacetimes \cite{CDL11} and improves the efficiency of inflation, with multiple accelerated expansions because 
$\G$ behaves as a further scalaron other than the Ricci scalar $R$ \cite{Paolella, Martino}.

In Ref.~\cite{CMM19fRG}, we considered a general $f(R,\G)$ theory
\begin{equation}
S = \frac{1}{2\kappa} \int d^4 x \sqrt{-g} f(R,\G) + S^{(m)}\,.
\end{equation}
The field equations are \cite{Atazadeh14}:
\begin{eqnarray}
&&R_{kl} -\tfrac{1}{2}  g_{kl} R= \kappa (\rho_m u_ku_l +p_m h_{kl} )\\
&&+( \nabla_k\nabla_l  -g_{kl}\square) f_R  + 2R (\nabla_k\nabla_l  - g_{kl}\square) f_\G -4(R_k{}^m\nabla_m\nabla_l +R_l{}^m\nabla_m\nabla_k)f_\G \nonumber\\
&& +4 (R_{kl}\square  + g_{kl} R^{pq}\nabla_p\nabla_q  +  R_{kpql}\nabla^p\nabla^q) f_\G     -\tfrac{1}{2}g_{kl}(R f_R+\G f_\G -  f)\nonumber \\
&& + (1-f_R) (R_{kl}-\tfrac{1}{2} g_{kl}R) \nonumber
\end{eqnarray}
where $f_R = \partial_R f $ and $f_{\G} =\partial_\G f $. The extra terms in the right hand side arise from geometry. Also in this case, we can apply the above scheme and derive a geometric perfect fluid. We can proceed as follows.

\noindent \textbf{Lemma} \\
In a four-dimensional FLRW spacetime, $\G$ is a perfect scalar.

 \noindent\textbf{Proof}  \\
In $d=4$: $R_{jklm}R^{jklm} = C_{jklm} C^{jklm} + 2 R_{jk}R^{jk} -\tfrac{1}{3}R^2$. In FLRW spacetimes, the Weyl tensor is zero,
then we obtain $\G =-2R_{jk}R^{jk} +\tfrac{2}{3}R^2$. According to Eq.\eqref{Ricci},  it is:
$\G = \tfrac{4}{3}\xi (R-2\xi) $. Since $\xi $ and $R$ are perfect scalars, also $\G$ is perfect. $\Box$
\\ 
Being $R$ and $\G$ perfect scalars, also $f_R$ and $f_{\G}$ are perfect scalars: $\nabla_k f_R = f_{RR} \nabla_k R + f_{R{\G}} \nabla_k {\G}=
-u_k  (f_{RR}\dot R +f_{R{\G}} \dot {\G})=-u^k\nabla_k f_R$ and similarly for $f_{\G}$. It turns out that the tensor in the r.h.s. of the field equations has the perfect fluid form and then can act as a geometric source leading the cosmic evolution \cite{Paolella}.

\section{Higher-order gravity}
Higher-order corrections can be invoked  in the effective action of gravitational interaction. This issue is mainly related to the renormalizability of gravity at ultraviolet scales but it can be considered also at cosmological scales \cite{Gottloeber}.

Higher-order gravity can be  characterized by the action
 \cite{Schmidt} \begin{equation}
S=\frac{1}{2\kappa} \int d^4 x \sqrt{-g} \, F(R,\square R, ...,\square^k R) + S^{(m)}\,,
\end{equation}
where $F$ is a twice differentiable function of $(k+1)$ real variables. The condition of minimum action yields the Einstein field equations, where the higher-order geometric terms are collected in the right-hand side as \cite{Schmidt}: 
\begin{eqnarray}
&&R_{ij}-\tfrac{1}{2}Rg_{ij} =\kappa T^{(m)}_{ij} + (1- \Theta_0) R_{ij} +\tfrac{1}{2}(F-R-2\square \Theta_0) g_{ij} 
+\nabla_i\nabla_j \Theta_0  \label{fieldeq}\\
&&-\tfrac{1}{2}\sum_{a=1}^k [g_{ij} \nabla_l
(\Theta_a\nabla^l (\square^{a-1} R)) -(\nabla_i\Theta_a) (\nabla_j \square^{a-1} R)  - (\nabla_j \Theta_a )( \nabla_i \square^{a-1} R)] \nonumber
\nonumber \\
&&\Theta_a = \sum_{b=a}^k \square^{b-a} \frac{\partial F}{\partial (\square^b R)}, \quad a=0,...,k.  \nonumber
\end{eqnarray}
Remarkably, high-order gravity has the property $\nabla^i T^{(HG)}_{ij} =0$.
In \cite{CMM20HO}, we proved that the correction terms are perfect fluid. In view of this statement, we have o prove the following\\

 \noindent\textbf{Lemma} \\
\textit{$F$ and $F^a=\partial F/\partial y_a$, as functions of $(R,\square R, ...,\square^k R)$, are perfect scalars.}
\\
\noindent \textbf{Proof}  \\
It is 
\begin{eqnarray*}
&&\nabla_j F = \sum_{a=0}^n \frac{\partial F}{\partial  y_a} \nabla_j (\square^a R) = -u_j \sum_{a=0}^n \frac{\partial F}{\partial  y_a} u^l\nabla_l (\square^a R) = -u_j (u^l\nabla_l F)\\
&& \nabla_j F^a = \sum_{b=1..k}  \frac{\partial F^a}{\partial y_b} \nabla_j (\square^b R)
= -u_j  \sum_{b=1..k}  \frac{\partial F^a}{\partial y_b} u^l\nabla_l (\square^b R) = -u_j (u^l\nabla_l F^a )
\end{eqnarray*} 
and then the statement is proven. $\Box$ 
\\ 
As a consequence, $\square^c F^a$ is perfect for any power $c$, as well as the linear combinations 
$\Theta_a =  \sum_{b=a}^k \square^{b-a} F^b$, with $a=0,1,...,k$. According to this result, perfect scalars can be defined also for higher-order derivatives.

\section{Quadratic gravity}
Quadratic gravity takes into account  corrections to the Hilbert-Einstein action where quadratic curvature invariants are considered \cite{DeserTekin03, Stelle}, that is 
\begin{equation}  
S=\int d^4 x \sqrt{-g} \Big[ \frac{1}{\kappa}(R-2\Lambda_0) + \alpha R^2 + \beta R_{ij} R^{ij}
 + \gamma \G \Big] + S^{(m)}  
 \end{equation}
where $\alpha$, $\beta$, $\gamma $, $\Lambda_0$ are parameters and $\G$ is the Gauss-Bonnet invariant \eqref{GB}.
Since here we restrict to dimension 4, the integral of the scalar $\G$ is surface term and then is zero. It can be ignored.
The variation of the action with respect to the metric gives the stress-energy tensor associated to the
quadratic gravity
\begin{eqnarray*}  
T^{(QG)}_{kl} &=& \frac{1}{\kappa} (R_{kl}- \tfrac{1}{2} R g_{kl} +\Lambda_0 g_{kl} ) +
2 \alpha R (R_{kl} -\tfrac{1}{4} R g_{kl}) +
(2\alpha + \beta) (g_{kl} \square - \nabla_k \nabla_l)R \nonumber \\
&&+\beta \square (R_{kl}-\tfrac{1}{2} R g_{kl} ) + 2\beta (R_{akbl}-\tfrac{1}{4} g_{kl} R_{ab} ) R^{ab}. \nonumber
\end{eqnarray*}
Despite the complicated expression,  the tensor has the perfect fluid form in FLRW in any dimension
\cite{CMM19fR}. According to the formalism of perfect scalars, the statement can be  easily checked considering the above expressions
\eqref{Riemann} for the Riemann tensor, and \eqref{Ricci} for the Ricci tensor. These expressions are functions of the perfect scalars $R$  and $\xi$.
\section{Conclusions}
From a cosmological and astrophysical points if view, the issues of dark energy and dark matter can be addressed if suitable perfect fluids, sourcing the Einstein field equations, are detected and related to some fundamental components.

The  difficulty to reveal particles beyond the Standard Model, despite the precision of running experiments,  forces to invoke new approaches to match the phenomenology. Clearly, galactic and cluster dynamics, from one side, and accelerated expansion of the Hubble flow, from the other side,  point out that problems of missing matter and exotic cosmic fluids have to be considered in some way.

Extensions of General Relativity seem  natural answers if geometric corrections behave as perfect fluids. 

Here, we demonstrated that, taking into account the concept of {\it perfect scalar}, it is possible to model out any extra contribution or  smooth function, entering the Hilbert-Einstein action, as perfect fluids. The result is naturally achieved in FLRW space-times once the Weyl tensor $C_{ijkm}$ is null and thanks to the special form of the Riemann and Ricci tensors that can be recast as function of perfect scalars $R$ and $\xi$. 

This feature allows to define effective equations of state $\displaystyle{w=\frac{p}{\rho}}$ in or out of the so called Zeldovich interval $0\leq w \leq 1$, valid for  standard matter perfect fluids.  As a consequence,  accelerated behaviors for the scale factor  $a(t)$ can be obtained both at   late  \cite{Capoz02} and early epochs  \cite{Starob80, Barrow88}. Specifically, we mean that the
presence of a perfect scalar in a   theory of gravity can also describe
perfect fluids with $w<0$, giving rise to models with accelerated behaviors like inflationary or dark energy dominated eras and, obviously,  a  state of no deceleration or
acceleration, with equation of state  $w=-1/3$,  is  also possible. The  equation of state for geometric perfect fluids depends on the parameters of the theory and allows  to suitably match cosmic evolution (see for example \cite{Hu}).

 It is worth noticing   that, in Ref.\cite{Gurses},  authors applied our methodology to the Einstein- Lovelock
theory of gravitation and found the energy density and isotropic pressure of the perfect fluid.
In general, the paradigm we traced here could be considered also for other formulations of gravity like teleparallel gravity, based on the torsion scalar $T$, or non-metric theories of gravity, based on non-metric scalar $Q$ \cite{Trinity, Jackson},  every  time a perfect scalar can be defined.

\section*{Acknowledgments}
SC  acknowledges the Istituto Nazionale di Fisica Nucleare (INFN) Sez. di Napoli (Iniziative Specifiche QGSKY and MOONLIGHT2) for the support.


%

\end{document}